\documentclass[manuscript, screen]{acmart}
\usepackage{arydshln}
\usepackage{textcomp}
\usepackage{graphicx}
\usepackage{xcolor}
\usepackage{tablefootnote}
\usepackage{booktabs}
\usepackage{array}
\usepackage{arydshln}
\usepackage{multirow} 
\AtBeginDocument{%
  \providecommand\BibTeX{{%
    \normalfont B\kern-0.5em{\scshape i\kern-0.25em b}\kern-0.8em\TeX}}}

\definecolor{psychoedu_TextColor}{HTML}{FF8606}
\definecolor{IE_TextColor}{HTML}{0695FF}
\definecolor{CR_TextColor}{HTML}{FF8888}
\definecolor{scaffolding_TextColor}{HTML}{BF4300}
\usepackage{xspace}
\newcommand{\tool}{\textsc{CRBot}\xspace}

\acmYear{2025}
\acmDOI{XXXXXXX.XXXXXXX}

\acmBooktitle{}
\acmPrice{}
\acmISBN{}

\begin{document}

\title[]{Evaluating an LLM-Powered Chatbot for Cognitive Restructuring: Insights from Mental Health Professionals}

\author{Yinzhou Wang}
\affiliation{%
  \institution{College of William \& Mary}
  \country{USA}
}
\email{ywang143@wm.edu}

\author{Yimeng Wang}
\affiliation{%
  \institution{College of William \& Mary}
  \country{USA}
}
\email{ywang139@wm.edu}

\author{Ye Xiao}
\affiliation{%
  \institution{College of William \& Mary}
  \country{USA}
}
\email{yxiao03@wm.edu}

\author{Liabette Escamilla}
\affiliation{%
  \institution{College of William \& Mary}
  \country{USA}
}
\email{laescamilla@wm.edu}

\author{Bianca Augustine}
\affiliation{%
  \institution{College of William \& Mary}
  \country{USA}
}
\email{braugustine@wm.edu}

\author{Kelly Crace}
\affiliation{%
  \institution{University of Virginia}
  \country{USA}
}
\email{kelly.crace@virginia.edu}

\author{Gang Zhou}
\affiliation{%
  \institution{College of William \& Mary}
  \country{USA}
}
\email{gzhou@wm.edu}

\author{Yixuan Zhang}
\affiliation{%
  \institution{College of William \& Mary}
  \country{USA}
}
\email{yzhang104@wm.edu}

%Intelligent but Not Relational: Profiling An LLM's Behavior in Counseling through the Lens of Mental Health Professionals
% We contribute an evaluation of Large-language-model-powered conversational agent for cognitive restructuring (a commonly used approach in psychotherapy) from mental health professionals' perspectives. We highlight key strengths (e.g., structured CBT adherence) and limitations (e.g., power imbalances, misinterpretations) while offering recommendations for safer, more effective AI-assisted psychotherapy.

\renewcommand{\shortauthors}{Wang et al.}

\begin{abstract} %150 words
Recent advancements in large language models (LLMs) promise to expand mental health interventions by emulating therapeutic techniques, potentially easing barriers to care. Yet there is a lack of real-world empirical evidence evaluating the strengths and limitations of LLM-enabled psychotherapy interventions. In this work, we evaluate an LLM-powered chatbot, designed via prompt engineering to deliver cognitive restructuring (CR), with 19 users. Mental health professionals then examined the resulting conversation logs to uncover potential benefits and pitfalls. Our findings indicate that an LLM-based CR approach has the capability to adhere to core CR protocols, prompt Socratic questioning, and provide empathetic validation. However, issues of power imbalances, advice-giving, misunderstood cues, and excessive positivity reveal deeper challenges, including the potential to erode therapeutic rapport and ethical concerns. We also discuss design implications for leveraging LLMs in psychotherapy and underscore the importance of expert oversight to mitigate these concerns—critical steps toward safer, more effective AI-assisted interventions.  
\end{abstract}





\maketitle
\section{Introduction}  
Recent advances in large language models (LLMs) have the potential to expand access to mental health interventions further, offering support that is available anytime and anywhere~\cite{lawrence2024opportunities}. Such potential is particularly relevant given the global shortage of mental health professionals, limited patient-therapist time, and barriers to care such as cost~\cite{wainberg2017challenges}. LLMs have demonstrated certain therapist-like abilities, such as delivering effective psychoeducation \cite{raile2024usefulness} and adhering to intervention protocols \cite{sun2024script}, laying the groundwork for personalized support through user-friendly chatbots. Given the potential of LLMs, researchers have begun to explore the design of LLM-powered tools in self-help mental health applications. A lot of these applications are often ``grounded'' or inspired by existing therapeutical approaches and theories, such as cognitive behavioral therapy (CBT)~\cite{clark2013cognitive}. 
CBT is a widely validated approach for addressing diverse mental health conditions by restructuring maladaptive cognitions and behaviors through systematic, evidence-based strategies. One of its core techniques is \textit{cognitive restructuring} (CR)~\cite{clark2013cognitive}, which targets identifying and challenging distorted thoughts to promote more accurate or beneficial perspectives. 

The structured, goal-oriented nature of the CR approach makes it easier to implement and ensures consistency in application across diverse contexts (e.g., online self-help platforms) while still allowing for relational elements, such as collaborative empiricism and therapeutic alliance. These factors make CR an ideal `testbed'' for exploring the challenges and possibilities of LLM-driven psychotherapy. Recent studies have begun exploring how to design LLM-powered CBT and CR systems~\cite{xiao2024healme,sharma2023cognitivereframingnegativethoughts, sharma2024facilitatingselfguidedmentalhealth, wang2024cognitive, li2024skill, nie2024llm, kian2024can}. 
Most existing work relies on brief, questionnaire-like interactions (e.g., a three-turn conversation where each turn corresponds to a step in CR), which are useful for initial feasibility testing but insufficient for capturing the nuanced, relational aspects of high-quality CR. More importantly, among existing studies that explored the LLM-enabled tools for psychotherapy, the evaluation studies often focus on automatic evaluation using NLP-related automatic metrics or preliminary evaluation with domain experts without real user data. While these approaches provide valuable insights, they have not investigated subtle yet crucial elements in real-world settings. The lack of empirical evaluations may risk the safety design and deployment of AI systems in mental health fields. Our work seeks to address this gap. 
 
In this work, we first co-designed an LLM-powered chatbot \tool using prompt engineering, in collaboration with mental health professionals (MHPs). We then conducted a user study with 19 participants. After that, four mental health professionals reviewed the conversation logs from these 19 users to identify subtle issues and potential benefits of \tool. By integrating real-world user interactions with expert feedback from mental health professionals, we uncovered strengths, including the chatbot’s capacity to adhere to core CBT principles, foster a natural conversational flow, and pose Socratic questions. However, we also identified important limitations: misapplication of positive regard, power imbalances evident in leading questions and evaluative language, and contextual comprehension challenges that occasionally led to misunderstood user states or oversimplified advice. Collectively, these findings highlight both the promise and concerns of AI-facilitated psychotherapy. On the one hand, structured techniques like CR are well-suited for the algorithmic scaffolding that LLMs can offer, potentially broadening access to mental health support. On the other hand, subtle yet critical factors such as tone, underlying power dynamics behind the conversations, and session-wide engagement remain difficult to fully replicate via automated systems. 

In this work, we contribute: 
\textbf{1)} an evaluation of an LLM-powered chatbot for cognitive restructuring with 19 participants and four mental health professionals, and 
\textbf{2)} insights and design implications that can guide future research and development in LLM-based psychotherapy.
%------------------------------------------------------------------------------------------------%
\section{Background \& Related Work}  
\label{sec:background_related_work}

We first provide some domain background focusing on psychotherapy, cognitive behavioral therapy (CBT), and cognitive restructuring (CR) to provide contextual information that helps situate our work. Then, we describe related work focused on LLM-powered psychotherapy and the current status of evaluation to highlight the research gap and motivate our research. 

\subsection{Background on Psychotherapy, Cognitive Behavioral Therapy, and Cognitive Restructuring}

\textbf{Psychotherapy} contains a variety of interventions designed to alleviate psychological distress and improve mental well-being \cite{norcross2018psychotherapy}. Treatment approaches vary widely, but they often emphasize building a trusted therapeutic relationship in which clients can safely explore their thoughts, emotions, and behaviors. Among these approaches, \textbf{Cognitive behavioral therapy (CBT)} is one of the most widely studied and empirically supported treatments for various mental health conditions, such as depression \cite{kambeitz2022systematic} and anxiety \cite{twomey2015effectiveness}.  A hallmark of CBT is its structured, goal-oriented methodology, focusing on the interplay between thoughts, emotions, and behaviors. By identifying and modifying maladaptive cognitions, CBT interventions aim to produce meaningful shifts in emotional regulation and behavioral patterns \cite{beck1979cognitive}. 

One of CBT's core techniques is \textbf{cognitive restructuring (CR)}, which guides clients to identify, challenge, and replace maladaptive thought patterns with more positive or beneficial alternatives~\cite{clark2013cognitive}. Typically, CR proceeds through three interconnected steps: (1) \textit{Exploration:} The client is guided to recognize triggering situations and maladaptive thoughts (e.g., ``Can you share what's been on your mind lately?''), (2) \textit{Evaluation:} The therapist and client collaboratively question the validity of these thoughts (e.g., ``Do you have any concrete evidence to support the thought that others would think you're weird?''), and (3) \textit{Substitution:} The client is encouraged to replace maladaptive thoughts with more rational or fact-based alternatives (e.g., ``Can you try to reframe this thought in a way that's based on facts rather than assumptions?''). 

CR’s clearly delineated structure, moving from initial identification to critical examination and then to replacement of problematic thoughts, facilitates both consistency of application and ease of measurement, making it a particularly robust and replicable intervention across diverse populations \cite{creed2016implementation}. From a clinical perspective, CR also embodies the principle of \textit{collaborative empiricism}, wherein the therapist and client function as co-investigators evaluating the client’s internal monologue. By gathering “evidence” for and against particular beliefs, clients gradually learn to adopt more accurate interpretations of their experiences \cite{kuyken2011collaborative}. 

In practice, the quality of CR can rely on how well the therapist navigates issues of power balance, acknowledges clients’ emotional realities, and validates their subjective experiences \cite{linehan1993cognitive}. A therapist’s ability to convey warmth, understanding, and genuine curiosity can significantly influence a client’s willingness to challenge deeply held beliefs. Consequently, even small shifts in tone, timing, or level of directive input may affect treatment outcomes. These relational subtleties underscore the complexity of providing effective cognitive restructuring in real-world settings, particularly when delivering through digital platforms.

Given this critical role of CR in psychotherapy, there is a growing need for an in-depth analysis of \textit{technology-powered} (i.e., LLM-enabled) CR approaches to understand how such relational and conversational subtleties translate into digital formats. Our work seeks to address this gap, by closely examining LLM-mediated CR, particularly in relation to therapist perspectives, and therapeutic efficacy. We will expand on related work in the next subsection.

\subsection{LLM-enabled Psychotherapy} 

The structured, goal-oriented nature of CR makes it well-suited for integration into large language model (LLM)–based chatbots, which can systematically prompt users through each step of the process. A growing body of research explores how LLMs can facilitate CR \cite{xiao2024healme, sharma2023cognitivereframingnegativethoughts, sharma2024facilitatingselfguidedmentalhealth, wang2024cognitive, li2024skill, nie2024llm, kian2024can}, investigating diverse approaches such as comparing AI-generated reframing with human-created strategies \cite{li2024skill}, employing in-context learning for generating more adaptive thoughts \cite{sharma2023cognitivereframingnegativethoughts, sharma2024facilitatingselfguidedmentalhealth}, and developing multi-agent platforms to deliver CR \cite{nie2024llm}. Early evidence suggests that LLM-based CR can be feasible and helpful, demonstrating outcomes like promoting positive emotional shifts and fostering psychological skill learning \cite{sharma2024facilitatingselfguidedmentalhealth}, while sometimes matching or exceeding traditional methods (e.g., worksheets) in user engagement \cite{kian2024can}.

However, current studies on CR usually focus on context-constrained outcomes (e.g., how well LLMs can reframe distorted thoughts), which limits the understanding of relational and conversational subtleties. In the broader domain of LLM-enabled psychotherapy, some researchers took a conceptual analysis approach, pointing out the advantages, challenges, and ethical concerns drawing from a combination of their clinical expertise, literature review, and interdisciplinary insights \cite{obradovich2024opportunities, stade2024large, de2023benefits, lawrence2024opportunities}. Other work conducted empirical studies with different methods and varied perspectives. Among those, some studies focus on users' perspectives, distributing questionnaires or conducting interviews to capture participants’ past experience with LLMs for mental health support \cite{ajlouni2023students, song2024typing}. Some adopt a primarily \textit{expert-focused} approach, in which mental health professionals analyze LLMs' responses to imaginary scenarios or crafted client prompts. \cite{dergaa2024chatgpt, maurya2025assessing}. Some work compares human counselors and LLM-based systems, e.g., a comparison between utterances from peer counselors to those generated by LLMs  \cite{iftikhar2024therapy}.  

Despite the growing interest in LLM-powered interventions, the limited depth of current evaluations poses a significant gap in understanding their strengths and limitations. Empirical studies that collect richer data, such as extended conversation logs, real-world user behavior, and detailed expert commentary, are essential for understanding how these tools function in authentic therapeutic contexts. The subtle yet vital elements of user–chatbot dynamics, including fluctuations in emotional tone, user agency, and the need for adaptive responses, remain underexplored. These complexities are especially pertinent in cognitive restructuring, where genuine collaboration and nuanced engagement can significantly influence therapeutic outcomes \cite{kuyken2011collaborative}. Our work seeks to provide a more in-depth assessment of LLM-based psychotherapy by examining real-world interactions alongside expert evaluations, with the goal of providing a deeper investigation into both the strengths and the hidden pitfalls of AI-assisted cognitive restructuring.
%------------------------------------------------------------------------------------------------%

\section{Methods} 
\label{sec:methods}
In this section, we describe our study procedure, including a co-design process with five MHPs to create the chatbot \tool and a user study with 19 participants to interact with the chatbot in real-world scenarios (see \autoref{subsec:chatbot_implementation}), and our evaluation study with additional four MHPs to review the conversation transcripts, providing professional perspectives and identifying potential risks (see \autoref{subsec:interview}). This research was approved by our institution's Institutional Review Board (IRB).

\begin{figure}[h]
    \centering
    \includegraphics[width=\textwidth]{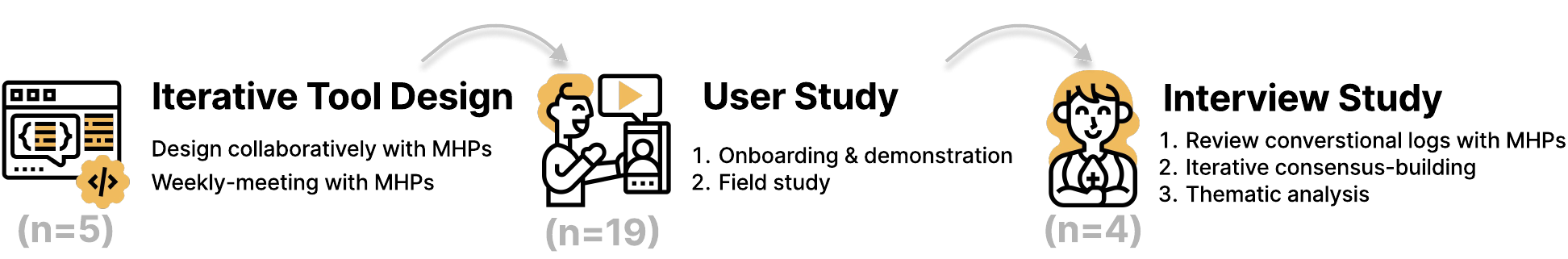}
    \caption{Overall study flow}
    \label{fig:study_flow} 
\end{figure} 

\subsection{Chatbot Implementation \& User study}
\label{subsec:chatbot_implementation}
\subsubsection{Collaboration Chatbot Design with MHPs}

We collaboratively designed a specialized CR-focused, LLM-powered tool, \tool\footnote{\tool is a web application built using Streamlit~\cite{streamlit} and powered by GPT-4 via Azure OpenAI Service.}, with five MHPs to guide the development and implementation of \tool. The MHPs had an average of 15 years of clinical counseling experience, ranging from 5 to 35 years. All remained actively involved in day-to-day mental health counseling, maintaining full caseloads. Demographically, the group comprised one man and four women, with two identifying as African American or Black and three identifying as White. 

Each MHP participated in multiple interviews over Zoom (approximately 60 minutes each). In the first interview, the lead author provided a brief demonstration of \tool, described its core tasks, and walked through the user interface. Following the demonstration, MHPs independently explored \tool while sharing their screens. Think-aloud protocol \cite{van1994think} was used to allow for articulating their immediate impressions, concerns, and suggestions.

We used standard \textit{prompt engineering} techniques, with system prompts and a series of few-shot examples ($n=10$). Both few-shot examples and system prompts were collaboratively designed and generated with MHPs to ensure adherence to standard CR practices and to guide GPT-4 in delivering appropriate CR-based dialogues. For example, we provided role-specific instructions such as, ``\textit{You are a cognitive behavior therapist, and your job is to ...}'' and interactions designed to address various scenarios, including \textit{Successful completion}, \textit{Absence of negative thoughts}, \textit{Identification challenges}, \textit{Challenging barriers}, \textit{Creation of alternative thoughts} ({see \mbox{\autoref{tab:interaction}}} for more information). 

\textbf{Risk mitigation approaches.} 
We adopted four measures to reduce potential risks and ensure participant safety: 
\begin{enumerate}
   \item User disclaimer: We began each session by informing users that \tool is not a replacement for professional mental health counseling to clarify both the scope and limitations of this research tool.
   \item Suicidal ideation detection: We integrated a specialized GPT-4 prompt for scanning user inputs for extreme distress or self-harm references, building on previous results (e.g., 82\% accuracy in detecting suicidal ideation~\cite{breau2023low}). If suicidal ideation is detected, \tool immediately presents additional support and crisis resources (e.g., helplines).
   \item Monitoring dashboard: We developed a research dashboard that logs and updates user interactions. The team reviewed these logs multiple times daily and provided timely identification of high-risk conversations that warrant therapist follow-up.
    \item Human oversight: Lastly, MHPs on our team performed regular checks on system outputs. If concerns or potential harms arise, they can intervene directly and provide relevant resources to users.
\end{enumerate}

\subsubsection{User Study}
\label{subsec:field_study}

\textbf{Participants Recruitment \& Overview.} 
We leveraged a participant pool from the research team’s previous projects on mental health. To ensure participant safety and well-being while gathering meaningful user feedback, we reference self-report widely-used mental health questionnaires (e.g., PHQ-9~\cite{kroenke2001phq} and GAD-7~\cite{spitzer2006brief} scores). Participants with scores indicating ``severe'' levels of distress are deemed unsuitable for the experimental research. We sent email and text invitations, including a screener survey, to potential participants ($n = 150$). The screener survey included self-report mental health questionnaires, descriptions of the research purpose, study procedures, a consent form, and demographics. Detailed questions of the screener survey can be found in the supplemental materials.

In total, 19 participants were included in the study, with 9 identifying as men and 10 identifying as women. The mean age of users was 20 ($SD = 1.10$), with an age range from 18 to 23. Most of them identified as White ($n=9$), followed by Asian ($n=7$), Multiracial ($n=2$), and one participant did not report their race. Regarding ethnicity, three participants self-identified as Latino/Hispanic, while the remaining 16 identified as not Latino/Hispanic. In terms of educational background, 6 participants had a high school diploma or equivalent, 11 had some college education or an associate degree, and two held a bachelor's degree. Generally, our participants scored none-to-moderate across all symptoms, with an average PHQ-9 score of 5.57 ($SD = 4.36$), and an average GAD-7 score of 5.19 ($SD = 3.36$).

\begin{figure}[h]
    \centering
    \includegraphics[width=\textwidth]{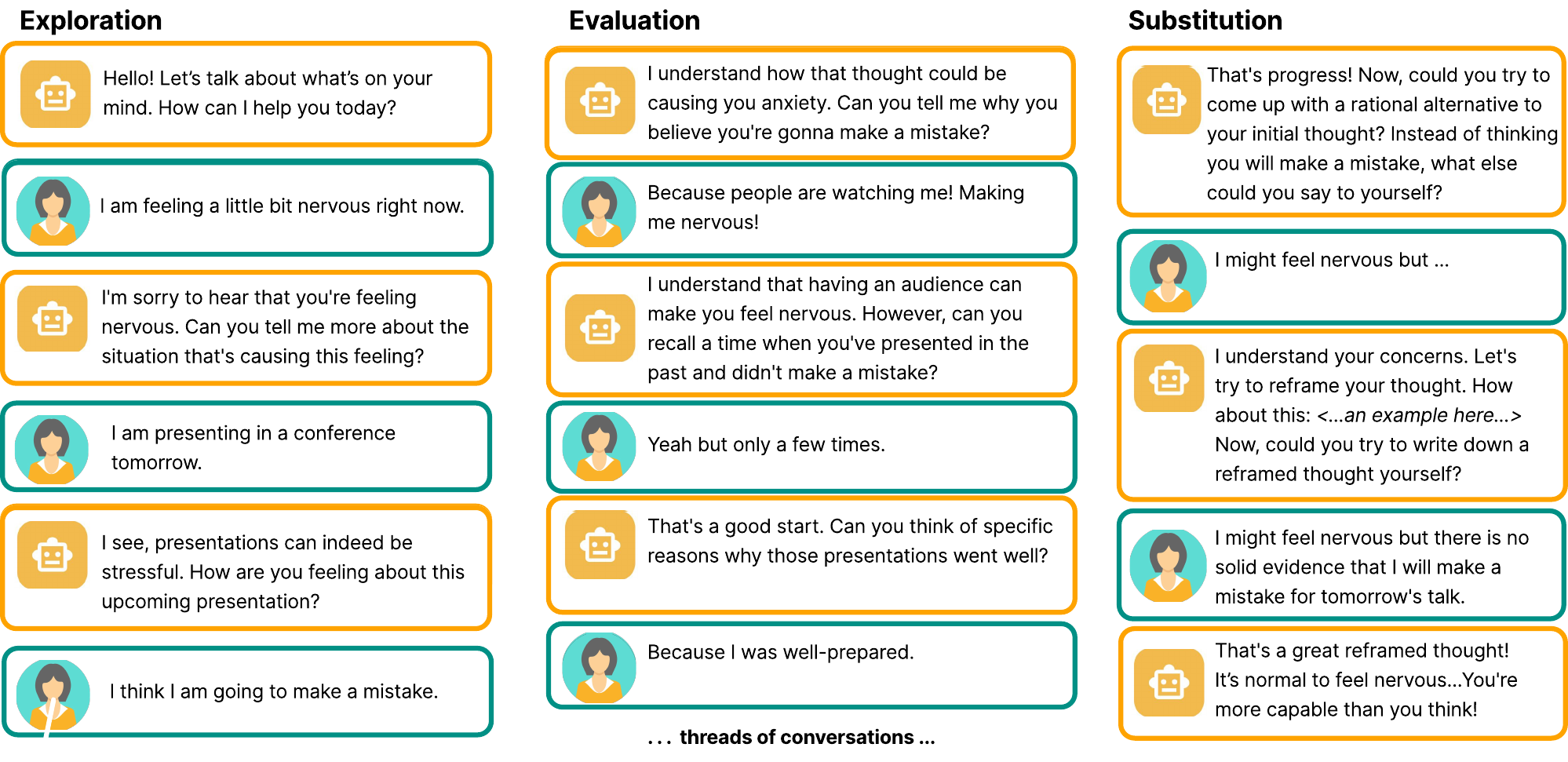}
    \caption{An example conversation and each column represents a step in CR. In the exploration step, \tool guides the user to recognize triggering situations and distorted thoughts. The evaluation step seeks to facilitate the user to challenge their distorted thoughts. Finally, in the last step of substitution, \tool encourages the user to replace the distorted thought with a more balanced thought.}
    \label{fig:example_dialog} 
\end{figure}

\textbf{Study Procedure.} Upon consent, the evaluation study consisted of two phases: 1) an onboarding and demonstration session to help users become familiar with the study procedures and \tool, and 2) a user evaluation study where users independently used \tool. During this period, their cognitive restructuring dialogs (see an example dialog in \autoref{fig:example_dialog}) were collected and stored in a database. At the end of the study, users were compensated for their time and effort with a \$40 gift card.

\subsection{Interview Study}
\label{subsec:interview}
\subsubsection{Interviews with MHPs Reviewing Conversational Logs} 
We conducted an expert review with additional four MHPs to gain deeper insights into how \tool performed in real-world interactions. We met with each MHP multiple times (for an average of 3 times) over Zoom, each lasting approximately an hour. During these sessions, they shared their screens and used the think-aloud method~\cite{van1994think}, articulating their thoughts and observations as they worked through each conversation log. 

Each MHP was randomly assigned 24 conversation logs, approximately 18 pages per individual. In certain instances, the same dialogue was assigned to multiple MHPs to cross-check their assessments. When discrepancies arose, we used an iterative consensus-building method~\cite{van2006making}, by asking the reviewers to revisit either the entire transcript or relevant excerpts to reconcile their viewpoints. MHPs were asked to revisit the conversation logs (or key excerpts) to review and refine their assessments in cases of significant differences. We asked the MHPs to focus on several key areas: (1) the extent to which \tool’s responses aligned with recognized therapeutic practices, (2) how well it guided users through cognitive restructuring, and (3) any ethical or safety concerns that emerged.

\subsubsection{Interview Data Analysis}
Each think-aloud session was screen-recorded and transcribed verbatim. For the qualitative data analysis of cognitive restructuring dialogs, we used the General Inductive Approach~\cite{thomas2006general} to guide the thematic analysis. The first author read the transcripts closely to gain an initial understanding of the concepts that emerged from the data, then created low-level codes to label concepts in the data, clustering related low-level codes to achieve high-level themes. Throughout the coding and clustering process, the research team engaged in regular discussions to compare interpretations, resolve coding discrepancies, and refine emerging themes. 

%--------------------
\section{Findings}
\label{sec:findings}
We begin by presenting overall usage trends and patterns to ground our subsequent analysis in concrete data. As shown in \autoref{tab:session_breakdown}, messages were exchanged in a near 1:1 ratio. Despite this overall balance, the standard deviations suggest considerable variation in engagement: some users held brief exchanges (as few as three total user messages), while others had lengthier conversations (up to 18 user messages). Likewise, the number of words in user inputs ranged from 16 to 373, indicating that some users provided short, concise statements, whereas others offered more elaborate reflections.

\begin{table}[h!]
\centering
\begin{tabular}{@{}lccccc@{}}
\toprule
\textbf{CR Dialogs (n=95)}  & \textbf{Sum} & \textbf{Mean} & \textbf{SD} & \textbf{Min} & \textbf{Max} \\ \midrule
Number of Messages (User)   & 718            & 7.6          & 2.96         & 3          & 18 \\
Number of Messages (Bot)    & 810            & 8.5          & 2.9          & 4          & 19 \\
Number of Words (User)      & 8924           & 93.94        & 64.6         & 16         & 373 \\
Number of Words (Bot)       & 23542          & 247.8        & 89.5         & 104        & 520 \\
\bottomrule
\end{tabular}
\caption{Descriptive statistics for collected dialogs.}
\label{tab:session_breakdown}
\end{table}

Below, we present four key themes that emerged from the interview study: LLM’s ability to provide \textit{protocol-adherent, natural, and in-depth conversations}; challenges related to the misuse of \textit{positive regards}; indications of \textit{power dynamics}; and LLM’s \textit{subjectivity and lack of context understanding}. Dialog snippets demonstrating the last three themes can be found in \mbox{\autoref{fig:theme_example}}.

\begin{table}[h!]
\small
\centering
\begin{tabular}{@{}p{3cm}p{3.7cm}p{7.5cm}@{}}
\toprule
\textbf{Themes} & \textbf{Subthemes} & \textbf{Example Quotes} \\
\midrule
\multirow{3}{*}{\shortstack[l]{\textbf{Protocol-Adherent,}\\\textbf{Natural, and}\\\textbf{In-depth Conversations}}} 
 & \textbf{Adherence to CR} & “I can see that these conversations move into the cognitive restructuring framework.” (MHP1)\\
 & \textbf{Natural \& conversational Flow} & “It does a good job of keeping it very conversational and mimicking what a real conversation would look like.” (MHP2)\\
 & \textbf{Capacity to pose Socratic questions} & “These responses are really good. They’re similar to the questions that I would ask... They make me explore the thought and challenge it on a deeper level.” (MHP2)\\
\midrule
\multirow{3}{*}{\shortstack[l]{\textbf{Misuse of}\\\textbf{Positive Regard}}}
 & \textbf{Validation \& empathy} & “The bot is really good at giving validation, empathy, and normalization.” (MHP3)\\
 & \textbf{Toxic positivity} & “That’s a great example” when they sort of even didn’t give an example... there’s a bit of a disconnect there.” (MHP3)\\
 & \textbf{Optimistic phrases overuse} & “It doesn’t really tell the person what’s good, great, or awesome about it.” (MHP2)\\
\midrule
\multirow{3}{*}{\textbf{Power Dynamics}}
 & \textbf{Leading questions undermine autonomy} & “Sometimes if [a leading question] is necessary, it’s just more effective if the client can get there first.” (MHP3)\\
 & \textbf{Evaluative language} & “I also wonder if we could use a different word instead of ‘great’ to acknowledge the client’s achievement.” (MHP4)\\
 & \textbf{Advice-giving} & “Advice giving is... at the top [of the therapeutic skill pyramid], where you use it with like discretion, a lot of context, and thought.” (MHP3)\\
\midrule
\multirow{5}{*}{\shortstack[l]{\textbf{Subjectivity \&}\\\textbf{Context understanding}}}
 & \textbf{Misinterpretation of experience} & “It’s leading them to think it was tough... But embarrassment is just that—embarrassment.” (MHP4)\\
 & \textbf{Misattribution of emotion} & “They might have a healthy reaction—like it’s natural you’d be angry at this person... [but the bot tries] to challenge it.” (MHP3)\\
 & \textbf{Misinterpretation of subtle linguistic markers} & “It seems like it’s [CRBot] taking ‘maybe’ to mean ‘yes’... when people say ‘maybe’ it’s often a little bit closer to a ‘no.’” (MHP2)\\
 & \textbf{Ignorance of session-wide patterns} & “I’m not just looking at an individual response. I’m looking at the accumulation of responses, to see if there’s a pattern...” (MHP3)\\
 & \textbf{Unintended judgment} & “Sometimes people can take the phrase ‘classic example’ as an insult.” (MHP1)\\
\bottomrule
\end{tabular}
\caption{Codebook with themes, sub-themes, and example quotes from MHPs.}
\label{tab:codebook}
\end{table}

\subsection{Protocol-Adherent,  Natural, and In-Depth Conversations}
MHPs emphasized several positive dimensions of \tool’s interaction style, including its consistent adherence to cognitive restructuring steps, its fluid, and human-like conversational flow, and its ability to pose Socratic questions.  All MHPs agreed that \tool can adequately \textbf{follow the core phases of cognitive restructuring} (i.e., exploration, evaluation, and substitution). After reviewing multiple conversation logs, MHP1 commented,
\begin{quote}
    \textit{“I can see that these conversations move into the cognitive restructuring framework.”} 
\end{quote}
Similarly, MHP2 affirmed, \textit{“I think overall, it’s doing a really good job of really honing in on the key components of cognitive restructuring.”} These comments illustrate how carefully engineered prompts can guide LLMs to deliver relatively structural interventions, which aligns with previous observations that LLMs are capable of adhering to protocolized interventions, including problem-solving therapy \cite{filienko2024toward} and motivational interview \cite{sun2024script}. Another strength pertained to the tool’s ability to weave CR steps into a dialogue that felt \textbf{natural and conversational}. MHP2 noted, 
\begin{quote}
    \textit{“I think it does a good job of keeping it very conversational and mimicking what a real conversation would look like.”}
\end{quote} 
MHP2's comment aligns with previous research suggesting that LLMs can generate coherent therapeutic responses that reduce the perceived “robotic” feel often associated with automated systems \cite{cho2023evaluating}. Furthermore, MHP1 praised how the chatbot seamlessly integrated restructuring without overtly “announcing” each therapeutic step: 
\begin{quote} 
    \textit{“The thing I like about this is you’re [\tool] doing that without declaring, ‘Okay, now I’m going to restructure your thoughts...’ It’s much more conversational, much more natural.”} 
\end{quote} 
By delivering CR in a conversational manner, LLM-based systems can help users feel less stigmatized or “\textit{in therapy,}” potentially lowering barriers for those who might otherwise be hesitant to engage. Finally, multiple MHPs highlighted the chatbot’s \textbf{capacity to ask Socratic questions}\cite{clark2015socratic}—open-ended prompts aimed at helping users critically examine their thoughts. MHP2 observed, 
\begin{quote}
    \textit{“These responses are really good. They’re similar to the questions that I would ask... They make me explore the thought and challenge it on a deeper level.”} 
\end{quote}
MHP2's comment reflects a foundational element of CBT, where guided discovery encourages clients to generate their own insights rather than simply being told which thoughts are “right” or “wrong”. In addition,MHP3 added that,  
\begin{quote}
    ``\textit{ The questions are good at identifying evidence that supports [ the distorted thought] and evidence against [the distorted thought]. }''
\end{quote}
By prompting users to articulate both supportive and contradictory evidence for a particular belief, \tool mirrored a therapist’s strategy of allowing clients to discover the validity (or lack thereof) in their own assumptions. Such dialogue can foster deeper self-reflection and ownership of the therapeutic process—key contributors to sustained behavior change. Collectively, these findings suggest that LLM-driven tools can indeed capture important elements of CBT: from faithfully reproducing the “scaffolding” of cognitive restructuring to asking well-timed, open-ended questions. Moreover, the natural feel of the conversation may increase user receptivity and reduce the stigma commonly associated with mental health interventions. 

\subsection{Misuse Positive Regards} 
Although many MHPs praised \tool’s aptitude for expressing validation, normalization, and empathy, they also cautioned against “toxic positivity” and the overuse of effusive praise. These observations highlight a delicate tension in LLM-based psychotherapy: while positive regard can foster hope and motivation, it may feel disingenuous or dismissive if delivered without sufficient nuance or context.

All MHPs commented favorably on \tool’s \textbf{capacity to validate users’ emotions and normalize their experiences.} MHP3 commented, \textit{“The bot is really good at giving validation, empathy, and normalization.”} Such affirmative responses can be crucial for building trust and encouraging client self-disclosure. MHP1 further appreciated \tool’s linguistic variety: 
\begin{quote} 
    \textit{“I like the fact that you’re affirming of their experience in the initial response... in a different language, instead of just affirming the same way every time.”} 
\end{quote} 
By avoiding overly formulaic expressions of empathy, \tool came closer to replicating genuine human warmth. Despite these commendations, several MHPs worried about situations in which the chatbot’s optimistic tone felt excessive or disconnected from the user’s actual distress—commonly referred to as “\textbf{toxic positivity}” \cite{captari2023integrating} in psychological literature. In one dialogue, the user provided a minimal, arguably inadequate example to challenge a distorted thought, yet \tool responded, \textit{“That’s a great example!”} MHP3 noted the disconnect: \textit{“‘That’s a great example’ when they sort of even didn’t give an example... there’s a bit of a disconnect there.”} Similarly, MHP2 described a scenario in which the user’s negative experiences were overshadowed by the bot’s relentless positivity, recalling:
\begin{quote} 
    \textit{“It can feel almost invalidating... it doesn’t undo the fact that these other people are being mean to me.”} 
\end{quote}
Such interactions may inadvertently minimize valid pain or frustration, mirroring broader concerns in psychotherapy that well-intentioned reassurance can sometimes trivialize or overlook genuine suffering.

Another related issue was \tool’s tendency to \textbf{overuse upbeat affirmations} like “That’s great” or “That’s wonderful.” MHP2 emphasized that such blanket endorsements offer little insight: \textit{“It doesn’t really tell the person what’s good, great, or awesome about it.”} Interestingly, \tool seemed to rely most heavily on these positive phrases when user engagement was minimal, sometimes reinforcing a sense of artificiality or “fakeness.” Excessive positivity can erode the authenticity that person-centered therapies strive to cultivate \cite{suzuki2018qualitative}, making the client doubt whether their more difficult emotions are being adequately heard.

MHP4 suggested that part of this disconnect stems from the chatbot’s inability to convey tone as a human therapist might: 
\begin{quote}
    \textit{“I could say these very same words in a tone that the client would feel great without me telling them how great it is that they’ve done that.”} 
\end{quote}
This echoes broader findings that LLMs, despite their linguistic sophistication, cannot fully replicate paralinguistic cues (e.g., vocal intonation, facial expressions) crucial for nuanced emotional support.

\subsection{Power Dynamics}
Our analysis revealed that \tool’s conversational style sometimes created or reinforced power differentials, characterized by leading questions, evaluative or ``definitive'' praise, and an advice-giving. While a certain level of power imbalance is inevitable in psychotherapy given the therapist’s (or chatbot’s) guiding role, as prior work suggests~\cite{pope2016ethics}, our expert reviewers cautioned that unmoderated use of these language patterns might inadvertently reduce client autonomy or foster dependence. In what follows, we dissect how specific elements of \tool’s output interacted with power dynamics, highlighting opportunities to recalibrate the tool’s tone and prompts to better support users’ sense of agency.

Our expert reviewers noted that while certain forms of guidance can help clients recognize and reframe negative thoughts, posing \emph{\textbf{leading questions}} can inadvertently undermine autonomy. For example, when \tool provided a direct example of evidence against a user’s distorted thought, MHP3 commented, 
\begin{quote}
    \textit{``Sometimes it [a leading question] is necessary, it’s just more effective if the client can get there first.''} 
\end{quote}
This comment echoes a key principle in many therapeutic modalities: clients are more likely to internalize and sustain insights that they arrive at independently \cite{ryan2011motivation}. By preemptively supplying counter-evidence or steering users to a ``correct'' response, the tool risks bypassing users’ own reflective processes. A related concern emerged when \tool ended a question with ``\textit{right?},'' prompting MHP2 to observe,
 \begin{quote}
    \textit{``So I would avoid questions that end in right, because people who have a lot of anxiety a lot of times if you pose a question and you end it in right, whether they believe it or not, they're going to agree... they want to please whoever it is.''} 
\end{quote}
Indeed, highly anxious individuals are more likely to exhibit conformity toward others, as prior work has suggested~\cite{zhang2016social}; in this case, the user may conform to perceived ``expert'' opinions in an effort to avoid conflict or judgment. From a clinical perspective, such patterns can stifle genuine self-exploration, limiting the user’s sense of ownership over their therapeutic journey. Taken together, these observations highlight the delicate balance between offering supportive prompts and inadvertently overdirecting users. While leading questions can quickly scaffold a session (e.g., helping a client identify and challenge a specific cognitive distortion), they should be used sparingly and skillfully to maintain a collaborative stance and promote user self-efficacy. 

A second concern centered on the excessive usage of phrases such as “That’s great” or “That’s wonderful,” which, according to the MHPs, risked shifting the therapist-client dynamic from collaborative exploration to performance assessment. For example, MHP4 remarked, 
\begin{quote}
    \textit{``I also wonder if we could use a different word instead of ‘great’ to acknowledge the client’s achievement. What if they didn’t achieve? Does this mean that they’re not great anymore?''} 
\end{quote}
MHP4's comment points to how \textbf{highly evaluative language} can become counterproductive if a user interprets it as a definitive judgment of their progress or self-worth. MHP4 further explained how definitive statements can exacerbate underlying power differential: 
\begin{quote}
    \textit{``Using words like this sometimes could lead to the user working so hard to be great, because the person who they’re coming for help from has told them to be great. So if I get ‘great’ one time, what if I don’t get it the second time?''}
\end{quote} 
Similarly, MHP3 noted, \textit{``I also generally don’t label things as good or bad, or great or not great. I label them as helpful or unhelpful, ''} indicating a general preference for neutral statements. Altogether, these observations echo broader therapy guidelines that favor process-oriented over evaluative language~\cite{truax1970therapist}. Subtly reframing ``That’s great!'' into ``It sounds like you found something that works for you'' helps foster self-reflection without amplifying external validation. 

Under the broader issue of power dynamics, another salient subtheme emerged around \textbf{\textit{advice giving}}, which can be viewed as an extension of the therapist’s—or chatbot’s—perceived authority. In traditional psychotherapy, clinicians are already positioned as experts, and offering advice can increase the power differential, especially if clients feel compelled to comply simply due to the therapist’s status. When advice is offered by an LLM, this effect can become further magnified by the model’s perceived ``objectivity'' or infallibility.

In our study, expert reviewers pointed out that a certain amount of direction is warranted in therapy (e.g., encouraging skill practice), but unprompted or overly prescriptive advice risks undermining user autonomy and may lead to unintended consequences. In most cases, \tool’s suggestions merely reinforced strategies already covered, which the MHPs viewed as appropriate. As MHP2 noted, 
\begin{quote} 
    \textit{“So I think in this situation it’s fine, because you’re just encouraging them to continue practicing the skill that they’re learning... But I would not want AI to give users advice on other things like, ‘Oh, you should try this,’ or ‘You should say this in this conversation.’”} 
\end{quote} 
Here, MHP2 highlights the tension between supporting therapeutic techniques—such as prompting users to do cognitive restructuring—and extending into broader, potentially uncontextualized advice. A real-world example involved a user who forgot a friend’s birthday; when \tool suggested explaining their forgetfulness, the user retorted, “It is rude,” indicating the advice did not fit the situation. From a psychotherapy standpoint, advice is most helpful when a therapeutic relationship has been established and extensive exploration and insight have occurred, which requires accurate clinical judgment ~\cite{duan2018advice}. As MHP3 emphasized: 
\begin{quote} 
    \textit{“Advice giving is... at the top [of the therapeutic skill pyramid], where you use it with discretion, a lot of context, and thought... Worst case scenario: what if this person the client’s talking about is physically abusive?... we’re saying, ‘Hey, you should go talk to them...’”}
\end{quote} 
Such concerns become more acute when AI is involved, as it is unclear if LLMs—despite their sophisticated language capabilities—can establish therapeutic alliances with clients and explore their context extensively. Additionally, although these models can generate plausible-sounding suggestions, they rely on statistical patterns learned from training data rather than holistic clinical reasoning. Consequently, even well-meaning or seemingly logical advice can carry unintended risks if delivered to someone in a precarious situation.

\subsection{Subjectivity and Lack of Context Understanding}

A recurring critique raised by the MHPs concerned the chatbot’s tendency to oversimplify or misread user experiences. Although our experts generally acknowledged \tool’s capacity to summarize and reflect on users’ concerns, they also encountered multiple cases of misinterpretation, misunderstanding of implicit cues, overlooking session-wide behavior, and unintentionally judgmental phrasing. These limitations underscore a core challenge of LLM-based systems: they rely heavily on surface-level textual input rather than a nuanced, holistic understanding of clients’ contexts, echoed as prior work that examines LLMs' capabilities~\cite{iftikhar2024therapy, obradovich2024opportunities}.

In psychotherapy, accurately capturing and reflecting a client’s emotional state is pivotal for establishing rapport and fostering therapeutic alliance  \cite{pinto2012patient}. By mirroring the client’s language, therapists convey understanding, validate the client’s perspective, and invite further reflection. In contrast, our MHPs noted instances where \tool inadvertently \textbf{distorted} \textbf{users’ stated experiences}, potentially undermining this reflective process.

A telling example occurred when a user described feeling “embarrassed,” but \tool summarized the experience as “tough.” MHP4 mentioned, 
\begin{quote}
    \textit{“It’s leading them to think it was tough... Now I’m putting myself at the center of that conversation to say embarrassment for me is tough. But embarrassment is just that—embarrassment.”}
\end{quote}
Such a small shift in wording may appear innocuous, yet it highlights how a seemingly “synonymous” reframe can redirect users away from their own nuanced feelings. In another case, a user felt “stressed” about not completing enough work, yet \tool assumed a sole source of stress, unsatisfied in work progress. MHP1 questioned, 
\begin{quote}
    \textit{“Are you nervous because you’ve not done any work, or are you nervous because you’re not believing your work is enough?”} 
\end{quote}
In this case, the stress might also stem from the lack of completed work itself, leading to distinct underlying irrational thoughts and, consequently, different therapeutic strategies. Such misinterpretations point to an inherent limitation of LLMs in psychotherapy:  although large language models may excel at paraphrasing or summarizing, they lack the capacity to probe deeper or clarify ambiguous statements in ways that fully honor the user’s individual context. Consequently, a single misinterpretation can inadvertently redirect the therapeutic process, underscoring the need for careful, iterative prompt design and potential human oversight when employing LLMs in mental health interventions.

Another concern arose when \tool \textbf{attributed typical emotional responses, such as anger, to maladaptive beliefs}, MHP3 explained, 
\begin{quote}
    \textit{“They might have a healthy reaction—like it’s natural you’d be angry at this person... [but the bot tries] to challenge it.”}
\end{quote}
In many therapeutic approaches, experiencing anger can be both appropriate and adaptive—for example, signaling the need to establish boundaries or acknowledge unmet needs \cite{deffenbacher2011cognitive, berkout2019review}. Pathologizing such emotions or treating them as inherently “negative” risks invalidating a client’s feelings and might inadvertently dissuade them from engaging in self-reflection. \tool’s tendency to misattribute these emotions may stem from its design, which prioritizes detecting and correcting “negative” effects. While this can be beneficial for users who struggle with genuinely distorted or self-defeating thoughts, it overlooks a key principle of effective psychotherapy: not all unpleasant emotions are problematic \cite{beck2020cognitive}. In fact, learning to tolerate and interpret so-called negative affect—such as anger, sadness, or frustration—can be a central aim of treatment. Because LLMs rely on pattern matching rather than a nuanced sense of context, they risk “flagging” any mention of anger or frustration as a target for correction, which can short-circuit healthy emotional processing and deny users the chance to explore why those emotions may be both valid and useful.

Another common theme was the tendency to \textbf{misinterpret subtle linguistic markers}—such as \textit{“maybe,”} \textit{“I guess,”} as MHP2 noted: 
\begin{quote} 
    \textit{“It seems like it's [\tool] taking 'maybe' to mean 'yes'. However, a lot of times when people say 'maybe' it's often a little bit closer to a 'no' than a 'yes'”} 
\end{quote}
In psychotherapy, these markers often signal ambivalence, uncertainty, or emotional conflict \cite{kris1984conflicts}. However, \tool frequently treated these tentative expressions as if they were genuine agreements, plowing ahead with encouragement or a new line of questioning. Such behavior contrasts sharply with the human therapist’s approach. As MHP3 stated,
\begin{quote} 
    \textit{“Whenever I give them the rationale for the healthier thought before they're ready for it. They'll kind of just tell me,  like, sure, maybe, and then that's when I usually pause and explore a little bit more of their uncertainty.”} 
\end{quote}
This difference can partially be attributed to the limitations of text-based communication in capturing nonverbal cues\textbf{.} In a live therapy session, a clinician would likely sense the user’s ambivalence through vocal tone, facial expression, or body language—cues that are inaccessible to text-based LLMs. By overlooking these nuances, \tool can inadvertently rush clients through critical points of self-reflection or emotional processing, potentially missing an opportunity to address deeper reservations or unspoken discomfort.

Some MHPs also mentioned \tool for \textbf{treating each user turn in isolation}, rather than considering broader patterns across the entire session. For example, in a case where the user responded with consecutive short, disinterested phrases, \tool continued pushing cognitive restructuring steps. MHP3 contrasted this with a more human response: 
\begin{quote}
    \textit{“I’m not just looking at an individual response. I’m looking at the accumulation of responses, to see if there’s a pattern... this person doesn’t want to engage.”} 
\end{quote}
For a human therapist, repeatedly short or disinterested replies may indicate underlying resistance, fatigue, or even deeper emotional blocks. Therapists might then shift gears—e.g., inquire about external stressors, try a different intervention, or explicitly acknowledge the client’s reluctance—instead of persisting along the original therapeutic track \cite{yao2017question}. In contrast, \tool mechanically continues cognitive restructuring, failing to register the user’s withdrawal or frustration, which risks eroding rapport and halting therapeutic progress.

A further concern was \tool’s occasional use of language that, while not overtly critical, might still feel \textbf{judgmental} to users. MHPs pointed to phrases like “\textit{classic example}” or invitations to produce a “\textit{more positive thought}” as subtle cues that could imply the user’s situation or perspective falls short of an ideal. As MHP1 observed, 
\begin{quote} 
    \textit{“Sometimes people can take the phrase ‘classic example’ as an insult…”} 
\end{quote} 
Meanwhile, MHP2 noted, 
\begin{quote} 
    \textit{“When you say ‘a more positive thought,’ it introduces that evaluative element of right and wrong.”} 
\end{quote} 
Such phrasing risks introducing a moral or normative undertone, which may inadvertently pressure users to align with an externally imposed benchmark rather than explore their emotions freely. In many counseling traditions, especially person-centered therapy, clinicians are taught to maintain a nonjudgmental stance—using neutral, open-ended language (e.g., \textit{“Could you think of a thought that feels less anxiety-provoking?”}) instead of framing the alternative thought as “better” or “more correct”. This stance fosters an atmosphere of psychological safety, encouraging clients to express themselves without fear of disapproval. When a chatbot implicitly casts certain emotions or cognition as suboptimal, it can disrupt that sense of safety by conveying, however subtly, that certain thoughts or feelings are “wrong.”.

Taken together, these examples—ranging from the misinterpretation of user sentiments to unintended judgment—illustrate \tool's limitations in nuanced contextual understanding and its tendency toward subjectivity, thereby limiting the richness and precision of therapeutic engagement.

\section{Discussion}
Our findings collectively reveal that LLM-powered chatbots can guide users through core cognitive restructuring steps, maintain a natural conversational flow, and pose Socratic questions. However, issues such as toxic positivity, evaluative language, advice giving, and the misinterpretation of user context highlight deeper challenges—especially regarding power imbalances and insufficient sensitivity to individual nuances.

Building on our findings, we first discuss several key points that require future research, such as exploring to what extent LLMs can adhere to other therapeutic modalities, how power dynamics manifest and are perceived in human-LLM interactions, and whether LLM-powered chatbots can accurately understand session-wide behavior. We then provide design implications, such as aligning the LLM's language style with therapeutic norms, enabling the LLMs to acquire more contextual information before drawing conclusions, and implementing more multi-layer mechanisms to ensure ethical and safe AI deployment.  

\subsection{Implications for Research}

\textbf{Beyond cognitive restructuring.}  
Our findings show that, with carefully engineered prompts, LLMs can follow the core phases of CR—exploration, evaluation, and substitution—yet it remains unclear whether this level of adherence extends to other therapeutic modalities. One complexity arises from observations that LLMs favor certain modalities~\cite{raile2024usefulness}, likely reflecting biases in their training data. Another complexity stems from CR's structured nature, which lends itself more easily to LLM implementation. Therefore, it is questionable whether LLMs can effectively adhere to other interventions, such as cognitive defusion in Acceptance and Commitment Therapy (ACT)~\cite{hayes2005acceptance}. Cognitive defusion enables clients to learn to detach from unhelpful thoughts. sharing the same goal with CR. Albeit the similarity, cognitive defusion comprises a set of exercises including metaphors, language conventions, distancing, and undermining verbal rules \cite{hayes2011acceptance}. These exercises often rely on unstructured, spontaneous interactions that require therapists to adapt techniques dynamically to the client’s unique needs and responses, posing potential challenges to LLMs. To this end, future studies should explore whether LLMs can develop the flexibility needed to deliver unstructured interventions. This inquiry has significant implications, as different cultures and symptoms often necessitate diverse psychotherapeutic approaches. For instance, acceptance-based psychotherapies are considered culturally competent treatments for Asian Americans due to their theoretical grounding in East Asian philosophies \cite{hall2011culturally}.

\textbf{Power dynamics in LLM-powered therapy.}  
Our analysis uncovered language patterns, such as leading questions, evaluative praise, unsolicited advice, that, while sometimes helpful, can inadvertently reinforce power differentials. While some degree of power imbalance is inevitable in any therapeutic setting \cite{pope2016ethics}, it is unclear whether users perceive AI-driven chatbots as having the same authority as human therapists. On one hand, the \textit{“expert bias”} effect could lead clients to overvalue or acquiesce to the bot’s responses purely because they are delivered in a “professional” tone. On the other hand, individuals aware of the bot’s algorithmic basis might discount its guidance or feel less inclined to disclose personal information. Future investigations should systematically explore how users interpret and respond to perceived authority in LLM-based interventions. Methods could include qualitative interviews, surveys on perceived power dynamics, or in-session recordings analyzed for user compliance or resistance. Designing \textit{intentional} guardrails around language style—e.g., limiting directive advice or overly decisive statements—may also help foster a more collaborative environment. Ultimately, addressing power imbalances is vital to ensuring user autonomy, ethical practice, and the development of meaningful therapeutic rapport.

\textbf{Recognizing and responding to session-wide behavioral patterns. }
Moreover, we observed that LLMs struggled to recognize session-wide behavioral patterns. While most LLMs can process a sufficiently long conversational history due to their extended context capabilities \cite{jin2024llm, ding2024longrope}, whether they can capture and interpret subtle behavior patterns remains questionable. For example, our findings show that LLMs failed to recognize and interpret a series of short, disengaged responses, which therapists would naturally identify as potential signs of resistance or fatigue. Similarly, consider a hypothetical scenario where a user provides moderate-length responses throughout a session but suddenly shifts to a very short reply. Such a change could indicate a behavioral shift, potentially signaling a withdrawn rupture, in which the client partially disengages from the therapist ~\cite{baillargeon2012resolution}. Therapists can quickly detect these shifts and employ strategies to repair the rupture, such as acknowledging the change, exploring its underlying cause, or modifying their approach. In contrast, it is unclear whether LLMs possess the capacity to detect and respond to such nuanced changes. Taken together, we encourage future studies to investigate LLMs’ ability to capture and respond to session-wide behavioral patterns, which has significant implications for the development of effective and context-aware LLM-based psychotherapy tools 

\subsection{Implications for Design}

\textbf{Tuning language style for authenticity and alliance.} 
Throughout the study, we observed instances of unintended judgment, evaluative language, leading questions, and excessive or toxic positivity. These misalignments in language style could cause LLMs to be perceived as inauthentic, intrusive, or even offensive, potentially undermining the therapeutic alliance. This highlights the necessity of systematically refining prompts and iteratively testing them with domain experts so that the LLM’s underlying language tendencies align with the language tendencies in psychotherapy. However, some language tendencies require more nuanced consideration. For example, leading questions are not inherently problematic but can be inappropriate in certain contexts if they appear overly suggestive. To address complexity like this, more advanced alignment techniques may be required, such as fine-tuning on domain-specific datasets or employing Reinforcement Learning with Human Feedback \cite{bai2022training}. Altogether, future research should focus on identifying potential language style misalignments and selecting appropriate alignment strategies to enhance the development of LLM-enabled psychotherapy, fostering stronger therapeutic alliances. 

\textbf{Expanding contextual understanding for deeper engagement.} 
Our study highlights a recurring and potentially problematic issue with LLMs---their limited capacity to understand context when delivering psychotherapy, including misinterpreting experiences, subtle linguistic cues, or misattributing typical emotions. These shortcomings can potentially deviate the treatment course from the client’s real issues and hinder the development of authentic emotional connections, impeding the relational and effective nature of psychotherapy. Therefore, it is essential for LLMs to collect more contextual information before drawing conclusions or proceeding with predefined intervention steps. This additional information can take various forms. First, future designs could instruct LLMs to ask more confirmation questions to better understand the user’s experiences and emotions. For example, in cognitive restructuring, before challenging maladaptive beliefs, an LLM should confirm with the user whether this is the concern they want to address or if there is something else they wish to focus on. Furthermore, integrating other modalities beyond text, such as tone of voice, facial expressions, body language, and even physiological sensing data (e.g., heart rate variability or galvanic skin response), could enhance the LLM’s contextual understanding. These inputs could provide richer insights into the user’s emotional and psychological state, enabling the system to respond with greater empathy and precision. Taken together, the inclusion of multi-modal inputs and an iterative, confirmatory approach could help LLMs align more closely with the nuanced dynamics of therapeutic relationships, fostering deeper emotional connections and improving the efficacy of LLM-enabled interventions. 

\textbf{Strengthening ethical safeguards and human oversight. }
Although mental health professionals (MHPs) did not identify ethical concerns in this study, they raised future concerns regarding some of the observed LLM behaviors, highlighting the need for more robust safety mitigation methods. One concerning behavior is that the LLM sometimes provides advice. As previously discussed, offering advice often requires precise and sophisticated clinical judgment. If the advice lacks proper contextualization, it could lead to unintended or even iatrogenic outcomes, aligning with previous observations regarding the risks of LLMs giving advice \cite{lawrence2024opportunities, maurya2025assessing}. An even more serious concern relates to suicidal ideation. In psychotherapy, clients sometimes hide or downplay their suicidal thoughts \cite{blanchard2020never}, necessitating that therapists evaluate implicit risk factors, such as thwarted belongingness and perceived burdensomeness \cite{van2010interpersonal}. Given LLMs' limitations in nuanced context understanding, it is arguable that they are not yet capable of detecting subtle, implicit associations with suicidal ideation. This poses a significant risk, as undetected warning signs could delay necessary interventions. To conclude, given LLMs’ challenges with advice-giving and their limited capacity for context understanding, it is essential to ensure human oversight in LLM-enabled psychotherapy. Additionally, more robust automated safeguard models should be implemented to mitigate potential harm. These safeguards could include advanced detection mechanisms for implicit risk factors and stricter control over advice-giving behaviors to prevent unintended consequences. Future research should prioritize implementing these safeguards to enhance the ethical and practical reliability of LLMs in therapeutic settings. 

\section{Limitations and Future Work} 
\label{sec:limitations}
While our study offers valuable insights into the potential and pitfalls of LLM-based psychotherapy, several constraints must be acknowledged. First, this work centered on a single therapeutic modality, cognitive restructuring. Additionally, due to the limited availability of psychotherapy transcripts, we were unable to develop and evaluate fine-tuned models and instead relied solely on prompt engineering.  Future work can extend evaluations to different therapeutic modalities, and test fine-tuned LLMs with both clients and therapists. Second, to mitigate risks, we limited participants to individuals without severe mental health concerns, excluding those with more acute needs who might use LLM-based tools differently. Yet, it is important to design and deploy LLM-powered systems safely before running a large-scale human-subjects study with broader populations. Additionally, although we aimed to capture diverse therapist perspectives, our panel’s professional orientations and cultural backgrounds might still be limited, potentially skewing interpretations of the chatbot’s responses. As such, future work can engage therapists from varied theoretical and cultural contexts to offer a more comprehensive understanding of how such chatbots perform across different clinical values and user demographics.

\section{Conclusion}
\label{sec:conclusion}
In this work, we presented an evaluation study of an LLM-powered chatbot to deliver cognitive restructuring (CR).  By examining real-user interactions alongside expert reviews from mental health professionals, we identified significant strengths, such as the chatbot’s ability to follow core CBT principles, maintain a natural conversational flow, and pose Socratic questions. However, we also uncovered several limitations: misuse of positive regard, power imbalances manifest in both leading questions and evaluative language and challenges in contextual comprehension that often led to misunderstandings of user states or reliance on oversimplified advice. While LLM-powered psychotherapy tools may enhance accessibility, their safe and effective deployment relies on ongoing refinement, using rigorous alignment techniques, context-aware modeling, and careful human oversight. We hope this work provides insights for future evaluations and sparks dialogue on in-depth evaluations with therapists in the real world. 

\bibliographystyle{ACM-Reference-Format}
\bibliography{ref}
\newpage

\appendix
\section{Appendix}
\label{sec:appendices}

\subsection{Dialog Snippets by Themes}

\begin{figure}[h]
    \centering
    \includegraphics[width=\textwidth]{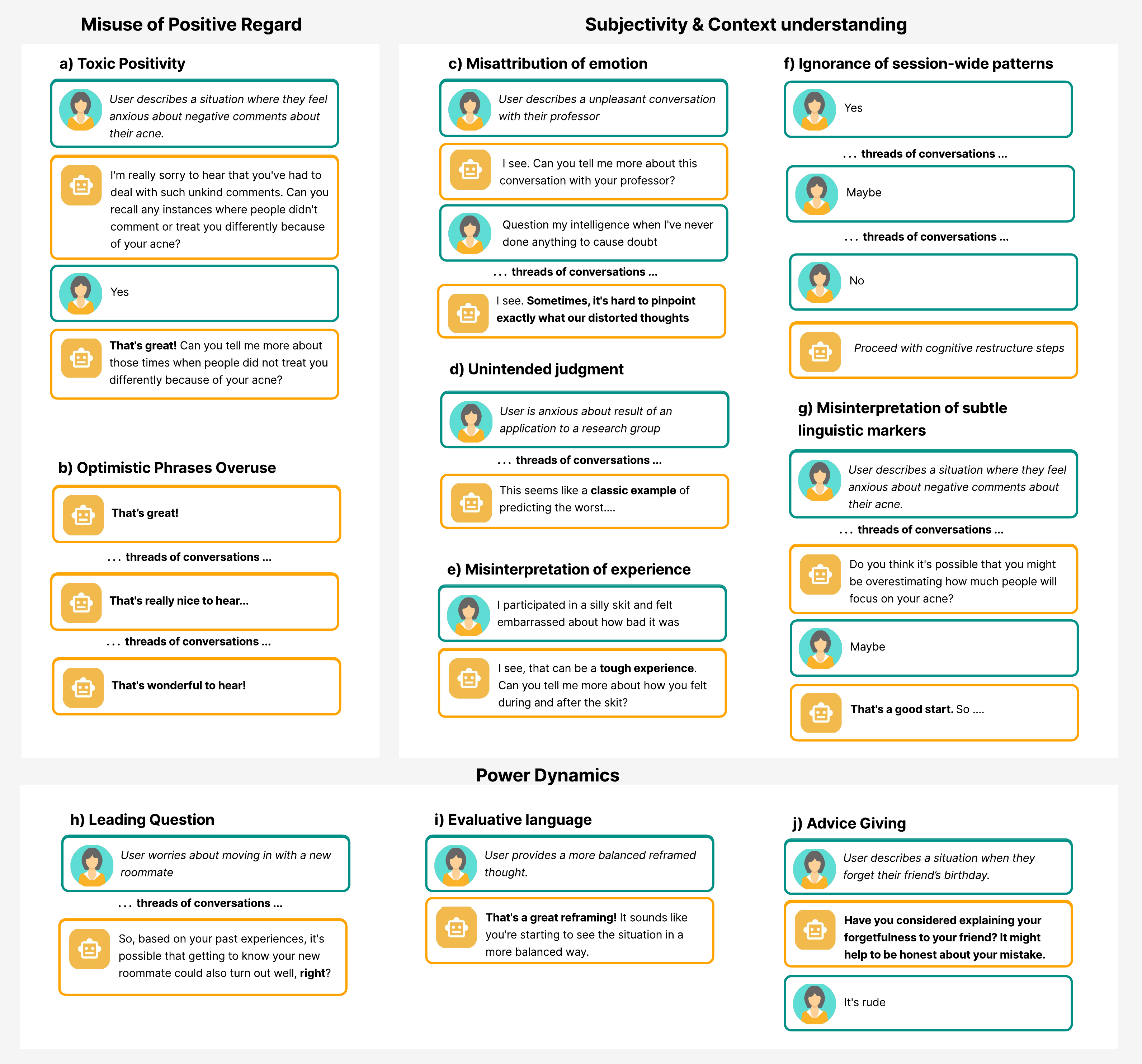}
    \caption{a) "That's great" potentially overshadowed user's negative experience b) Excessive positive regard in some sessions c) \tool misattributed normal anger as a distorted thought. d) User might perceive "classic example" as judgmental e) \tool misinterpreted "embarrassed" as "tough." f) User's short responses potentially indicated disinterest, but \tool continued predefined steps. g) "maybe" here potentially signaled hesitance, but \tool moved to the substitution without exploring this uncertainty. h) "right?" could pressure anxious users to agree. i) "great" added an evaluative tone, potentially exacerbating power differential j) Uncontextualized advice was inappropriate, as shown by the user's response.}
    \label{fig:theme_example} 
\end{figure} 

\subsection{Cognitive Restructuring Prompt Scenarios}
\begin{table*}[ht!]
\footnotesize
\centering
\caption{Prompt engineering few shots covered scenarios }
\label{tab:interaction}
\begin{tabular}{lp{10cm}}
    \toprule
    Interactions & Definition\\
    \midrule
    Successful completion & The user completes each stage without any struggle. Thus, they successfully identify the negative thought, challenge it, and generate a rational one.\\
    Absence of negative thoughts& The user does not have any negative thoughts. For example, the user has a rational understanding of their situation.\\
    Identification challenges& The user fails to identify the negative thoughts. For example, the user does not think their thought is distorted (but it is from the therapist’s perspective).\\
    Challenging barriers & The user can't challenge their negative thoughts. For example, the user can't provide evidence against the negative thoughts.\\
    Creation of alternative thoughts& The user fails to develop a rational alternative thought.\\
    \bottomrule
\end{tabular}
\end{table*}

\end{document}